\documentclass[aps,prd,onecolumn,preprintnumbers,notitlepage,superscriptaddress,article
nofootinbib,amsfont,amssymb,10pt,singlespacing] {revtex4-1}
\usepackage{amsmath,bm}
\usepackage{times}
\usepackage{braket}
\usepackage{color,graphicx}
\usepackage{slashed}
\usepackage{hyperref}
\usepackage{graphics}
\usepackage{subfigure}
\usepackage{multirow,makecell}
\usepackage{textcomp}
\usepackage{mathtools}
\usepackage{float}
\usepackage{siunitx}
\usepackage{amssymb}
\usepackage{graphicx}

\newcommand{\beq}{\begin{eqnarray}}
\newcommand{\eeq}{\end{eqnarray}}

\def\lsim{ {\ \lower-1.2pt\vbox{\hbox{\rlap{$<$}\lower6pt\vbox{\hbox{$\sim$}
}}}\ } }
\def\gsim{ {\ \lower-1.2pt\vbox{\hbox{\rlap{$>$}\lower6pt\vbox{\hbox{$\sim$}
}}}\ } }
\definecolor{Red}{rgb}{1.,0.,0.}

\definecolor{Blue}{rgb}{0.,0.,1.}

\definecolor{nicered}{rgb}{0.7,0.1,0.1}
\definecolor{nicegreen}{rgb}{0.1,0.5,0.1}

\begin{document}

    \bibliographystyle{apsrev}

    \hypersetup{colorlinks,citecolor=nicegreen,linkcolor=nicered}

	
	\title{Study of $B_{c} \to J/\psi (a_1(1260)$, $b_1(1235)$, $a_2(1320)$, $K_2^*(1430)) $  decay with a perturbative QCD approach}
	
	\author{Yun~Zhao}
	\email[Electronic address:]{2956368954@qq.com}
	\affiliation{School of Physical Science and Technology,
		Southwest University, Chongqing 400715, China}

	\author{Xian-Qiao~Yu}
	\email[Electronic address:]{yuxq@swu.edu.cn}
	\affiliation{School of Physical Science and Technology,
		Southwest University, Chongqing 400715, China}
	
	\date{\today}
	
	\begin{abstract}
		
	Motivated by systematic observations of $ B_c $ decays at LHCb, we analyze $B_c^+ \to J/\psi (A, T)$ decays (where $A$ denotes axial-vector mesons $a_1(1260)^+$, $b_1(1235)^+$, and $T$ denotes tensor mesons $a_2(1320)^+$, $K_2^*(1430)^+$) within the perturbative QCD (pQCD) factorization framework. The framework employs light-cone distribution amplitudes (LCDAs) to regulate nonperturbative dynamics,  utilizing Sudakov suppression to control endpoint divergences. Our predictions indicate branching ratios of $10^{-3}\sim 10^{-2}$ for $B_c^{+} \to J/\psi A$ and $10^{-5}\sim 10^{-4}$ for $B_c^{+} \to J/\psi T$ decays, which seem to be in the reach of future experiments. We also provide predictions for the polarization fractions of these decays. Based on quark helicity analysis, the longitudinal polarization amplitudes are expected to dominate the branching ratios.
	\end{abstract}

	\maketitle
	
	%
	%

	\section{Introduction}\label{sec:intro}
	
The $B_c$ meson was first discovered experimentally in 1998 by the CDF Collaboration in proton-antiproton ($p\bar{p}$) collisions at a center-of-mass energy of $\sqrt{s} = 1.8$~TeV at the Fermilab Tevatron collider \cite{Abe:1998fb}. This landmark discovery was made through the analysis of the semileptonic decay channel $B_c \to J/\psi \ell \nu$ (where $J/\psi$ is the charmonium state $c\bar{c}$, and $\ell$ denotes a lepton such as an electron or muon). A distinctive feature of this decay process is the reconstruction of $J/\psi$ via its dilepton final states ($\mu^+\mu^-$ or $e^+e^-$), while the lepton $\ell$ and neutrino $\nu$ provide key signatures of weak interactions. Up to now, the LHCb collaboration has experimentally established a comprehensive set of $B_{c}$ meson decay channels, such as $B_c^+ \to J/\psi \ell^+ \nu_\ell$ ( where $\ell = e$, $\mu$, $\tau$) \cite{Abulencia:2006,Abazov:2009}, $B_c^+ \to \chi_{c0} \pi^+$ \cite{Aaij:2016},$B_c^+ \to J/\psi \pi^+$ \cite{Aaij:2012}, $B_c^+ \to J/\psi D_{s}^{(*)+}$ \cite{Aaij:2013}, $B_c^+ \to \psi(2S) \pi^+$ \cite{Aaij:2013a},   $B_c^+ \to J/\psi  \pi^+ \pi^- \pi^+$ \cite{Aaij:2012a}, $B_c^+ \to J/\psi K^+$ \cite{Aaij:2013ab}, $B_c^+ \to J/\psi K^+ K^- \pi^+$ \cite{Aaij:2013abc}, and $B_c^+ \to B_s^0 \pi^+$ \cite{Aaij:2013abcd} for the first time. At the LHC with a proton-proton collision energy of $\sqrt{s} = 14~\si{\tera\electronvolt}$, the inclusive production cross-section of $B_c$ mesons has a theoretical prediction of $\SI{1}{\micro\barn}$. With an integrated luminosity of $\SI{1}{\femto\barn^{-1}}$, this corresponds to the production of $\sim \num{1e9}$ $B_c$ mesons \cite{Gao:2010} before detector acceptance and branching fraction corrections, providing a robust statistical sample for precision studies of their properties. Leveraging the LHCb experiment's high-precision detection capabilities and future upgrades for the High-Luminosity LHC (HL-LHC), $B_{c}$ physics will remain pivotal in probing strong interactions, weak decay dynamics, and potential beyond-Standard-Model phenomena.
	
Compared to the $B_{u}$, $B_{d}$, and $B_{s}$ mesons, the $B_{c}$ meson stands out as a unique probe in particle physics. Composed of two heavy quarks with distinct flavors ($\bar{b}$ and $c$), the $B_{c}$ meson decays exclusively through weak interactions, with strong and electromagnetic annihilation processes being kinematically forbidden. The presence of two heavy constituents allows for a distinctive decay mechanism: when one quark decays, the other acts as a passive spectator. This dual-heavy nature results in a significantly shorter lifetime for the $B_{c}$ meson compared to other b-flavored mesons \cite{Aaij:2013a}, underscoring the pivotal role of the ${c}$ quark in its decay dynamics. With its diverse decay pathways, the $B_{c}$ meson serves as an exceptional laboratory for investigating non-leptonic weak decays of heavy mesons, probing the limits of the Standard Model, and uncovering potential signatures of new physics beyond current theories \cite{Brambilla:2004}.
	
 Various theoretical approaches have been utilized to explore decays of $B_{c}$ mesons to charmonium states, including pQCD\cite{Li:1995}, generalized factorization \cite{Hsiao:2017}, QCD factorization \cite{Chang:2018}, QCD sum rules \cite{Kiselev:2000}, the Bethe-Salpeter equation \cite{Chang:1994}, relativistic quark models \cite{Ivanov:2006}, non-relativistic QCD (NRQCD)\cite{Faustov:2022}, and the covariant light-front approach \cite{Cheng:2004,Wang:2009a,Wang:2009b,Wang:2007}. In the $B_c$ meson system, the applicability of the pQCD approach originates from its distinct double heavy-flavor structure, which establishes a clear hierarchy of hard energy scales. Although the $B_c^+$ meson consists of a bottom quark and a charm antiquark ($\bar{b}c$), with the charm quark mass being significant, the ratio $m_c / m_{B_c} \sim  0.2$ indicates that the dominant energy carrier is the much heavier b quark. This justifies approximating the $B_c$ as a heavy-light system. Consequently, the $k_T$-factorization theorem, successfully applied to $B_u$ and $B_d$ meson decays, can be extended to b-quark decays within the $B_c$ framework.
The specific decay process involves three characteristic scales: M(the $B_c$ mass), $m_2$(the charmonium mass), and $\bar{\Lambda}$, which characterizes the non-perturbative binding energy. These scales obey the hierarchy $\bar{\Lambda} \ll m_2 \ll M$, enabling a consistent systematic power expansion in $m_2 / M$ and $\bar{\Lambda} / m_2$.
Within the heavy-quark and large-recoil limits, and based on the $k_T$-factorization theorem, the decay amplitude factorizes into a convolution of a hard kernel with the meson wave functions. The hard kernel is calculable perturbatively at the leading order in the $\alpha_s$ expansion, corresponding to single-gluon exchange. Higher-order radiative corrections generate single logarithmic divergences, which are absorbed into the meson wave functions. The double logarithmic divergences from overlapping collinear and soft divergences are summed to all orders into a Sudakov factor, effectively suppressing long-distance interactions. Non-perturbative effects, such as final-state interactions, are systematically incorporated into the universal wave functions of the initial and final state mesons. Although these wave functions are not calculable from first principles, their universality allows them to be constrained from other experimental processes and utilized as non-perturbative inputs.
Therefore, the existence of hard scales ensures the validity of the perturbative expansion. The combination of the factorization theorem, the Sudakov mechanism, and the parameterization of wave functions establishes a comprehensive pQCD framework for systematically handling non-perturbative effects, including final-state interactions, in $B_c$ meson decays.
\setlength{\parskip}{0pt}  

Up to now, the two-body hadronic decays of $B_c$ mesons have been extensively studied in theoretical literature \cite{Kiselev:2000,Chang:1994,AbdElHady:2000,Colangelo:2000,Ebert:2003,Fu:2011,Naimuddin:2012,Kar:2013,Qiao:2014,Ke:2014,Ebert:2010,Hernandez:2006,Ivanov:2005,Ivanov:2006,Wang:2008,Chang:2001,Kiselev:2002,Wang:2012,Dhir:2013,Wang:2009b,Zhu:2017,Rui:2014,Rui:2015,Bondar:2005}. The decays of $B_c$ mesons into final states that include a charmonium meson hold distinctive research importance. Primarily, such decay channels offer a sensitive probe for investigating strong interaction dynamics within heavy meson systems. Furthermore, these processes entail two distinct energy scales characterized by the bottom quark mass $m_b$ and the charm quark mass $m_c$. Within the framework of quantum chromodynamics (QCD), the calculation of higher-order corrections is formulated in terms of a systematic expansion in the mass ratio $m_c/m_b$. This approach differs from the standard perturbative expansion that uses $\Lambda_{\mathrm{QCD}}/m_b$ as the small parameter (where $\Lambda_{\mathrm{QCD}}$ denotes the QCD scale). Given that $m_c/m_b$ is not negligibly small, this expansion may induce substantial correction effects, thus necessitating more refined theoretical treatment. This fundamental observation requires a theoretical framework distinct from that used for ordinary $B$ mesons. Specifically, the work addresses this by adopting a non-relativistic $B_c$ wave function, which inherently incorporates the double-heavy mass effects, providing a more physical starting point than LCDAs for light mesons.
Additionally, by reconstructing the properties of charmonium states through their decays to known final states, these $B_c$ decays offer a direct probe into the characteristics of charmonium. In this work, we calculate the decay amplitudes and branching fractions for $B_c^+ \to J/\psi (A, T)$, where $A$ denotes axial-vector mesons ($a_1(1260)^+$ and $b_1(1235)^+$) and $T$ represents tensor mesons ($a_2(1320)^+$ and $K_2^*(1430)^+$). For simplicity, the following abbreviations will be used throughout this work unless otherwise specified: ${a_1}$ denotes the axial-vector meson $a_1(1260)$ , ${b_1}$ denotes the axial-vector meson $b_1(1235)$ , ${a_2}$ denotes the tensor meson $a_2(1320)$ , and ${K_2^*}$ denotes the tensor meson $K_2^*(1430)$. The axial-vector mesons $a_1$ and $b_1$ correspond to $^3P_1$ and $^1P_1$ states respectively, which cannot mix in the SU(3) limit due to G-parity constraints. While $^3P_1$ mesons have symmetric LCDAs similar to vector mesons \cite{Cheng:2007,Cheng:2008}, $^1P_1$ states exhibit antisymmetric amplitudes with vanishing decay constants in this limit. For tensor mesons with quantum numbers $J^{P} = 2^{+}$, a factorizable amplitude proportional to the matrix element $\langle T \vert j_{\mu} \vert 0 \rangle$, where the current $j_{\mu}$ is of the $(V \pm A)$ or $(S \pm P)$ type, does not contribute, as this matrix element vanishes due to constraints from Lorentz covariance\cite{Zou:2013,Cheng:2010}. This makes tensor meson emission strictly forbidden in naive factorization, explaining why such predictions significantly underestimate experimental results and highlighting the importance of non-factorizable diagrams and annihilation processes. We extend previous pQCD analyses of $B_c \to J/\psi a_1(1260)$ \cite{Najafi2015} by incorporating several improvements: (1) updating CKM matrix elements \cite{ParticleDataGroup:2024} and hadronic parameters ; (2) including intrinsic $b$-dependence in the $B_c$ meson wave function to better account for transverse momentum effects \cite{Rui:2014}; (3) expanding the study to previously unexplored channels including $B_c \to J/\psi b_1$, $a_2$, and ${K_2^*}$  decays within the pQCD framework. These specific decay channels have also been investigated in a recent study\cite{Liu:2023bc} using a pQCD-based approach. However, our work differs fundamentally in the treatment of charm-quark mass effects and the wave function model for the $B_c$ meson. The cited study introduces a modified Sudakov factor, $s_c(Q, b)$, into the traditional pQCD framework, aiming to explicitly include the charm quark mass $m_c$ in the resummation of soft gluon radiation.
In contrast, our analysis employs the conventional pQCD approach with a distinct strategy to address the $m_c/m_b$ challenge. We use a non-relativistic approximation for the $B_c$ light-cone wave function through a delta function $\delta(x \, - \, m_c/M_{B_c})$, which fixes the longitudinal momentum fraction of the charm quark. This approach naturally incorporates the non-relativistic nature of the heavy-heavy $B_c$ system and reduces sensitivity to endpoint uncertainties in the light meson distribution amplitudes.
Thus, while both studies describe the same physical processes, their main difference lies in the theoretical treatment of heavy-quark dynamics: the improved pQCD modifies the evolution kernel via a new Sudakov factor, while our approach optimizes the initial-state wave function within the standard pQCD formalism. These complementary approaches provide valuable insights for both experimental tests at LHCb and Belle II and further understanding of non-perturbative QCD effects.

This work is organized as follows: Section \ref{sec:pert} outlines the theoretical foundation of the pQCD framework and describes the essential wave functions employed in the computation of $B_{c} \rightarrow \  J/\psi(A,T) $ decay amplitudes. Section \ref{sec:numer} provides the numerical outcomes and analyses. Section \ref{sec:summary} summarizes the core findings of this study. For completeness, the full explicit expressions of all helicity amplitudes are compiled in the Appendix.

	%
	%
	\section{The Theoretical Framework and Helicity Amplitudes}\label{sec:pert}
	
	\subsection{The Wave Functions}
	
	The effective Hamiltonian of the $B_{c}$ non-leptonic decays into charmonium and axial-vector mesons or tensor mesons is given by \cite{Buchalla:1996}
	
	\begin{equation}
	\mathcal{H}_{\text{eff}} = \frac{G_F}{\sqrt{2}} V_{cb}^* V_{uq} \left[ C_1(\mu) O_1(\mu) + C_2(\mu) O_2(\mu) \right],\label{eq:hamiltonian}
	\end{equation}
	where $ q \in \{s, d\} $ denotes a light down-type quark. $ G_F $ is the Fermi constant, and $ V_{cb}^* $, $ V_{uq} $ represent CKM matrix elements. The Wilson coefficients $ C_{1,2}(\mu) $ (encoding short-distance effects above scale $ \mu $) and their corresponding four-quark operators $ O_{1,2}(\mu) $ are defined as:
	
	\begin{equation*}
	\begin{split}
	&O_1(\mu)= \bar{b}_\alpha \gamma^\nu (1 - \gamma_5) c_\beta \otimes \bar{u}_\beta \gamma_\nu (1 - \gamma_5) q_\alpha, \\
	&O_2(\mu)= \bar{b}_\alpha \gamma^\nu (1 - \gamma_5) c_\alpha \otimes \bar{u}_\beta \gamma_\nu (1 - \gamma_5) q_\beta.
	\end{split}
	\end{equation*}
	where $\alpha$ and $\beta$ are color indices (summation over repeated indices implied). Since the Hamiltonian involves four distinct quark flavors, these decays are free from penguin operator contamination, hence direct CP violation is naturally absent.
	
The calculation is performed in the $B_c$ meson rest frame. The momenta are defined in light-cone coordinates as follows: ${P_1}$ for the $B_c$ meson, ${P_2}$ for the recoiling charmonium, and ${P_3}$ for the emitted light meson.
	\begin{equation}
	\begin{split}	
	&\emph{P}_{1} = \frac{M}{\sqrt{2}}(1, 1, \mathbf{0}_T) ,  \\
	&\emph{P}_{2} = \frac{M}{\sqrt{2}}\left(1 - r_3^2, r_2^2, \mathbf{0}_T\right) ,  \\
	&\emph{P}_{3} = \frac{M}{\sqrt{2}}\left(r_3^2, 1 - r_2^2, \mathbf{0}_T\right).
	\end{split}
	\end{equation}
The mass ratio $r_{2,3}$ is defined as $ m_{2,3}/M$, where $M$ ($m_2$) denotes the mass of the $B_c$ meson ( $J/\psi$ meson), while $m_3$ represents the mass of the axial-vector (tensor) meson. The momentum of the valence quarks $k_{1,2,3}$ is represented by the parameterization:
	
	\begin{equation}
	\begin{split}	
	&k_{1}= x_1 P_1 + \mathbf{k}_{1T}, \\
	&k_{2}= x_2 P_2 + \mathbf{k}_{2T}, \\
	&k_{3}= x_3 P_3 + \mathbf{k}_{3T},
	\end{split}
	\end{equation}
	where parameters $k_{iT}$ and $x_i$ are defined as the transverse momentum and longitudinal momentum fraction, respectively, of the quark/antiquark confined within the meson. When the final state includes a vector charmonium in conjunction with an axial-vector meson, the formalism employs the longitudinal polarization vector $\epsilon_L$ and the transverse polarization vector $\epsilon_T$, which serve to describe the polarization states as follows:
	\begin{equation}
	\begin{split}
	\epsilon_{2L} &= \frac{1}{\sqrt{2(1 - r_3^2)} \ r_2} \left(1 - r_3^2, -r_2^2, \mathbf{0}_T\right), \quad \epsilon_{2T} = \left(0, 0, \mathbf{1}_T\right);\\
	\epsilon_{3L} &= \frac{1}{\sqrt{2(1 - r_2^2)} \ r_3} \left(-r_3^2, 1 - r_2^2, \mathbf{0}_T\right), \quad \epsilon_{3T} = \left(0, 0, \mathbf{1}_T\right).
	\end{split}
	\end{equation}
	The polarization vectors satisfy the normalization conditions $\epsilon_{L}^{2} = -1, \epsilon_{T}^{2} = 1 $ and the orthogonality relations $ \epsilon_{2L} \cdot P_2 = \epsilon_{3L} \cdot P_3 = 0 $.
	
	For the spin-2 polarization tensor $\epsilon^{\mu\nu}(\lambda)$ satisfying the transverse condition $\epsilon_{\mu\nu}p_2^\nu = 0$ \cite{Cheng:2010,Cheng2011}, its helicity components  $\lambda = 0, \pm1, \pm2$ can be constructed from the polarization vectors $\epsilon$ of vector mesons as follows:
	\begin{equation}
	\begin{split}
	&\epsilon^{\mu\nu}(\pm2)= \epsilon^\mu(\pm1)\epsilon^\nu(\pm1), \\
	&\epsilon^{\mu\nu}(\pm1)= \frac{1}{\sqrt{2}} \left( \epsilon^\mu(\pm1)\epsilon^\nu(0) + \epsilon^\mu(0)\epsilon^\nu(\pm1) \right), \\
	&\epsilon^{\mu\nu}(0)= \frac{1}{\sqrt{6}} \left( \epsilon^\mu(+1)\epsilon^\nu(-1) + \epsilon^\mu(-1)\epsilon^\nu(+1) \right) + \sqrt{\frac{2}{3}} \epsilon^\mu(0)\epsilon^\nu(0).
	\end{split}
	\end{equation}
	where $\epsilon(\pm1)=\epsilon_{3T}$ represents the transverse polarization vector, and $\epsilon(0)=\epsilon_{3L}$ represents the longitudinal polarization vector. For clarity and computational efficiency in subsequent calculations, we introduce another polarization vector $
	\epsilon_{\bullet \mu}(\lambda)=m_{T} \frac{\epsilon_{\mu \nu(\lambda)} v^{\nu}}{P_{3} \cdot v}
	$ \cite{Wang2011}, which satisfy
	\begin{equation} \label{polarization vector}
	\begin{split}
	\epsilon_{\bullet\mu}(\pm 2) &= 0, \\
	\epsilon_{\bullet\mu}(\pm 1)&=m_{T}\sqrt{\frac{1}{2}}\frac{\epsilon_{3 L}\cdot v}{P_{3}\cdot v}\epsilon_{3 T\mu}, \\
	\epsilon_{\bullet\mu}(0)&=m_{T}\sqrt{\frac{2}{3}}\frac{\epsilon_{3 L}\cdot v}{P_{3}\cdot v}\epsilon_{3 L\mu}.
	\end{split}
	\end{equation}
	
In this work, within the non-relativistic framework, we employ the $B_c$ meson distribution amplitude defined in Ref. \cite{Zhou2012},
	\begin{equation}
	\Phi_{B_{c}}(x) = \frac{i f_{B}}{4N_{c}} \left[ \left( \not p + M_{B_{c}} \right) \gamma_{5} \delta \left( x - \frac{m_{c}}{M_{B_{c}}} \right) \right] \exp \left( -\frac{\omega_{B}^{2} b^{2}}{2} \right)  \label{PhiBc}
	\end{equation}
	with $\omega_B$ being the shape parameter, $N_c$ the color number, and the last exponent term representing the $k_T$ dependence. In our calculations, we adopt the shape parameter $\omega_B = (0.60 \pm 0.05)~\text{GeV}$~\cite{Zhou2012}, consistent with our previous analyses of double-charm decays of the $B_c$ meson.
	
For $J/\psi$ charmonium states, wave functions have been comprehensively studied within the NRQCD framework. The quantum wave functions of the vector $J/\psi$ meson are characterized by distinct longitudinal ($\Phi^L$) and transverse ($\Phi^T$) components, expressed as~\cite{Rui:2014}
	\begin{align}
	\Phi_{J/\psi}^{L}(x) = \dfrac{1}{\sqrt{2N_{c}}} \left\{ m_{J/\psi} \notin_{L} \phi^{L}(x) + \notin_{L} \not P_{2} \phi^{t}(x) \right\}_{\alpha\beta}, \\
	\Phi_{J/\psi}^{T}(x) = \dfrac{1}{\sqrt{2N_{c}}} \left\{ m_{J/\psi} \notin_{T} \phi^{V}(x) + \notin_{T} \not P_{2} \phi^{T}(x) \right\}_{\alpha\beta}
    \end{align}
here, $\epsilon_{L}$ and $\epsilon_{T}$ are the longitudinal and transverse polarization vectors of the $J/\psi$ meson, respectively; $\phi^{L}(x)$ and $\phi^{T}(x)$ are the twist-2 distribution amplitudes, while $\phi^{t}(x)$ and $\phi^{V}(x)$ correspond to the twist-3 distribution amplitudes, which can be expressed as \cite{Li2006}:
	\begin{align}
		\phi^{L}(x) = \phi^{T}(x) &= 9.58\,\frac{f_{J/\psi}}{2\sqrt{2N_{c}}}\,x(1-x)\left[\frac{x(1-x)}{1-2.8x(1-x)}\right]^{0.7}, \\
		\phi^{t}(x) &= 10.94\,\frac{f_{J/\psi}}{2\sqrt{2N_{c}}}\,(1-2x)^{2}\left[\frac{x(1-x)}{1-2.8x(1-x)}\right]^{0.7}, \\
		\phi^{V}(x) &= 1.67\,\frac{f_{J/\psi}}{2\sqrt{2N_{c}}}\,\bigl[1+(2x-1)^{2}\bigr]\left[\frac{x(1-x)}{1-2.8x(1-x)}\right]^{0.7}.
	\end{align}
	
	For axial-vector mesons (\(a_1\) and \(b_1\)), their wave functions involve one longitudinal polarization state (L) and two transverse polarization states (T). These can be expressed as \cite{Liu2012}:
\begin{align}
	\Phi_{A}^{L}(x) = &\frac{1}{\sqrt{6}}\gamma_{5}\left\{m_{A} \not{\epsilon}_{A}^{*L} \phi_{A}(x) + \not{\epsilon}_{A}^{*L}\not P \phi_{A}^{t}(x)+ m_{A} \phi_{A}^{s}(x)\right\}_{\alpha \beta}, \\
	\Phi_{A}^{T}(x) = &\frac{1}{\sqrt{6}} \gamma_{5}\left\{m_{A} \not{\epsilon}_{A}^{*T} \phi_{A}^{v}(x) + \not{\epsilon}_{A}^{*T} \not P \phi_{A}^{T}(x)+ m_{A} i \epsilon_{\mu \nu \rho \sigma} \gamma_{5} \gamma^{\mu} \epsilon_{T}^{* \nu} n^{\rho} v^{\sigma} \phi_{A}^{a}(x)\right\}_{\alpha \beta}
\end{align}
	where \(x\)  represents the light-cone momentum fraction of a valence quark within the meson, with  \(n = (1, 0, \mathbf{0}_T)\) and \(v = (0, 1, \mathbf{0}_T)\) being dimensionless light-like basis vectors. Here we adopt the convention \(\epsilon^{0123} = 1\) for the Levi-Civita tensor.
	
	The leading-twist (twist-2) distribution amplitudes for axial-vector mesons with orbital configuration \(^3P_1\) and \(^1P_1\) can be expressed through distinct functional forms for longitudinal($J_z = 0$) and transverse($J_z = \pm 1$) polarization states \cite{Liu2012}:
	\begin{align}
	\phi_{A}(x) &= \frac{3 f_{A}}{\sqrt{2 N_{c}}} x(1-x) \left[ a_{0A}^{\parallel} + 3 a_{1A}^{\parallel}(2x-1)+ a_{2A}^{\parallel} \frac{3}{2} \left(5(2x-1)^{2} - 1\right) \right] \\
	\phi_{A}^{T}(x) &= \frac{3 f_{A}}{\sqrt{2 N_{c}}} x(1-x) \left[ a_{0A}^{\perp} + 3 a_{1A}^{\perp}(2x-1)  + a_{2A}^{\perp} \frac{3}{2} \left(5(2x-1)^{2} - 1\right) \right]
	\end{align}
	where $f_A$ is the decay constant. Here, the definitions of these distribution amplitudes satisfy the following relationships:
	\begin{equation}
	\begin{aligned}
	\int_{0}^{1} \phi_{{}^{3}\!P_{1}}(x)  dx &= \frac{f_{{}^{3}\!P_{1}}}{2\sqrt{2N_{c}}},
	& \quad \int_{0}^{1} \phi_{{}^{3}\!P_{1}}^{T}(x)  dx &= a_{0,{}^{3}\!P_{1}}^{\perp} \frac{f_{{}^{3}\!P_{1}}}{2\sqrt{2N_{c}}}; \\
	\int_{0}^{1} \phi_{{}^{1}\!P_{1}}(x)  dx &= a_{0,{}^{1}\!P_{1}}^{\parallel} \frac{f_{{}^{1}\!P_{1}}}{2\sqrt{2N_{c}}},
	& \quad \int_{0}^{1} \phi_{{}^{1}\!P_{1}}^{T}(x)  dx &= \frac{f_{{}^{1}\!P_{1}}}{2\sqrt{2N_{c}}}.
	\end{aligned}
	\end{equation}
	
	For axial vector mesons, the twist-3 distribution amplitudes adopt these functional forms \cite{Liu2012}:
	\begin{equation}
	\begin{split} \label{distribution A}
	\phi_{A}^{t}(x) &= \frac{3 f_{A}}{2\sqrt{2 N_{c}}}\left\{a_{0A}^{\perp}(2 x-1)^{2} + \frac{1}{2} a_{1A}^{\perp}(2 x-1)\left(3(2 x-1)^{2}-1\right)\right\}, \\
	\phi_{A}^{s}(x) &= \frac{3 f_{A}}{2\sqrt{2 N_{c}}}\frac{d}{dx}\left\{x(1-x)\left(a_{0A}^{\perp} + a_{1A}^{\perp}(2 x-1)\right)\right\};  \\
	\phi_{A}^{v}(x) &= \frac{3 f_{A}}{4\sqrt{2 N_{c}}}\left\{\frac{1}{2} a_{0A}^{\parallel}\left(1+(2x-1)^{2}\right) + a_{1A}^{\parallel}(2x-1)^{3}\right\},  \\
	\phi_{A}^{a}(x) &= \frac{3 f_{A}}{4\sqrt{2 N_{c}}}\frac{d}{dx}\left\{x(1-x)\left(a_{0A}^{\parallel} + a_{1A}^{\parallel}(2x-1)\right)\right\}.
	\end{split}
	\end{equation}
	here, \( x \) denotes the quark's momentum fraction. The Gegenbauer moments \( a_{i,A}^{\parallel(\perp)} \) have been extensively studied in the literature; in this work, we adopt the following values \cite{Yang2007}:
	\begin{equation}
	\begin{split}
	a_{0,^{3}P_{1}}^{\parallel} & = 1,  \quad a_{0,^{1}P_{1}}^{\perp} = 1;\\
	a_{0,b_{1}}^{\parallel} &= 0.0028\pm 0.0026, \quad a_{0,a_{1}}^{\perp}= 0; \\
	a_{1,b_{1}}^{\parallel} &= -1.95 \pm 0.35,\quad a_{1,a_{1}}^{\perp}= -1.04 \pm 0.34; \\
	a_{2,a_{1}}^{\parallel} &= -0.02 \pm 0.02, \quad a_{2,b_{1}}^{\perp}= 0.03 \pm 0.19.
	\end{split}
	\end{equation}
	
The intrinsic $b$ dependence is incorporated in the heavy-meson wave function $\phi_B$ but excluded for the light axial-vector meson $\phi_A$. Given the demonstrated insignificance of such dependence in light mesons, we preliminarily assume it is similarly nonessential for the currently undetermined $a_1$ and $b_1$ wave functions.

For light tensor mesons with $J^{P} = 2^{+}$, the five polarization states ($\lambda = \pm2, \pm1, 0$) are described by a rank-2 tensor $\epsilon_{\mu\nu}^{(\lambda)}$, where angular momentum conservation restricts decays to only the longitudinal and transverse polarizations ($\lambda = 0, \pm1$), while forbidding the maximal helicity states ($\lambda = \pm2$). Consequently, the generic tensor meson wave function can be defined as \cite{Zou:2013}:
\begin{equation}
\begin{split}
\Phi_{T}^{L} &= \frac{1}{\sqrt{6}}\left[m_{T}\not \epsilon_{\bullet L}^{*} \phi_{T}(x) + \not \epsilon_{\bullet L}^{*}\not P \phi_{T}^{t}(x) + m_{T}^{2} \frac{\epsilon_{\bullet} \cdot v}{P \cdot v} \phi_{T}^{s}(x)\right]  \\
\Phi_{T}^{\perp} &= \frac{1}{\sqrt{6}}\left[m_{T} \not\epsilon_{\bullet \perp}^{*} \phi_{T}^{v}(x) + \not \epsilon_{\bullet \perp }^{*} \not P \phi_{T}^{T}(x) + m_{T} i \epsilon_{\mu \nu \rho \sigma} \gamma_{5} \gamma^{\mu} \not \epsilon_{\bullet \perp}^{* \nu} n^{\rho} v^{\sigma} \phi_{T}^{a}(x)\right]
\end{split}
\end{equation}
where the distribution amplitudes have the following explicit forms \cite{Wang2011}:
\begin{equation}
\begin{split}
\phi_{T}(x) &= \frac{f_{T}}{2\sqrt{2 N_{c}}}\phi_{\|}(x),
\quad \phi_{T}^{t}(x) = \frac{f_{T}^{T}}{2\sqrt{2 N_{c}}} h_{\|}^{(t)}(x), \\
\phi_{T}^{s}(x) &= \frac{f_{T}^{T}}{4\sqrt{2 N_{c}}}\frac{d}{d x}h_{\|}^{(s)}(x),
\quad \phi_{T}^{T}(x) = \frac{f_{T}^{T}}{2\sqrt{2 N_{c}}} \phi_{\perp}(x), \\
\phi_{T}^{v}(x) &= \frac{f_{T}}{2\sqrt{2 N_{c}}} g_{\perp}^{(v)}(x),
\quad \phi_{T}^{a}(x) = \frac{f_{T}}{8\sqrt{2 N_{c}}} \frac{d}{d x} g_{\perp}^{(a)}(x).
\end{split}
\end{equation}

The twist-2 LCDA can be expanded in Gegenbauer polynomials $C_n^{3/2}(2x-1)$ with Gegenbauer moments as coefficients. In particular, its asymptotic form is:
\begin{equation}
\phi_{\|,\perp}(x) = 30x(1-x)(2x-1)
\end{equation}

Using the QCD equations of motion, the two-particle twist-3 distribution amplitudes can be expressed in terms of both twist-2 LCDAs and three-particle twist-3 LCDAs. In the Wandzura-Wilczek approximation (where three-particle terms are neglected), the asymptotic forms of twist-3 LCDAs are derived as \cite{Wang2011}:
\begin{equation}
\begin{split}
h_{\|}^{(t)}(x) &= \frac{15}{2}(2x-1)\left(1 - 6x + 6x^{2}\right), \\
h_{\|}^{(s)}(x) &= 15x(1-x)(2x-1), \\
g_{\perp}^{(a)}(x) &= 20x(1-x)(2x-1), \quad
g_{\perp}^{(v)}(x) = 5(2x-1)^{3}.
\end{split}
\end{equation}

\subsection{Helicity Amplitudes}

Based on the effective Hamiltonian in Eq.~\eqref{eq:hamiltonian}, the decay $B_c \to J/\psi(a_1,b_1)$ can be described by the four types of Feynman diagrams shown in Fig.~\ref{fig:fig1}. These diagrams give rise to the factorizable emission amplitude $\mathcal{F}_{e}$ and the nonfactorizable emission amplitude $\mathcal{M}_{e}$. Since the operators $\mathcal{O}_{1,2}$ are of $(V-A)(V-A)$ current type, these diagrams collectively yield the total decay amplitude:

\begin{figure}
	\centering
	\includegraphics[width=0.5\linewidth]{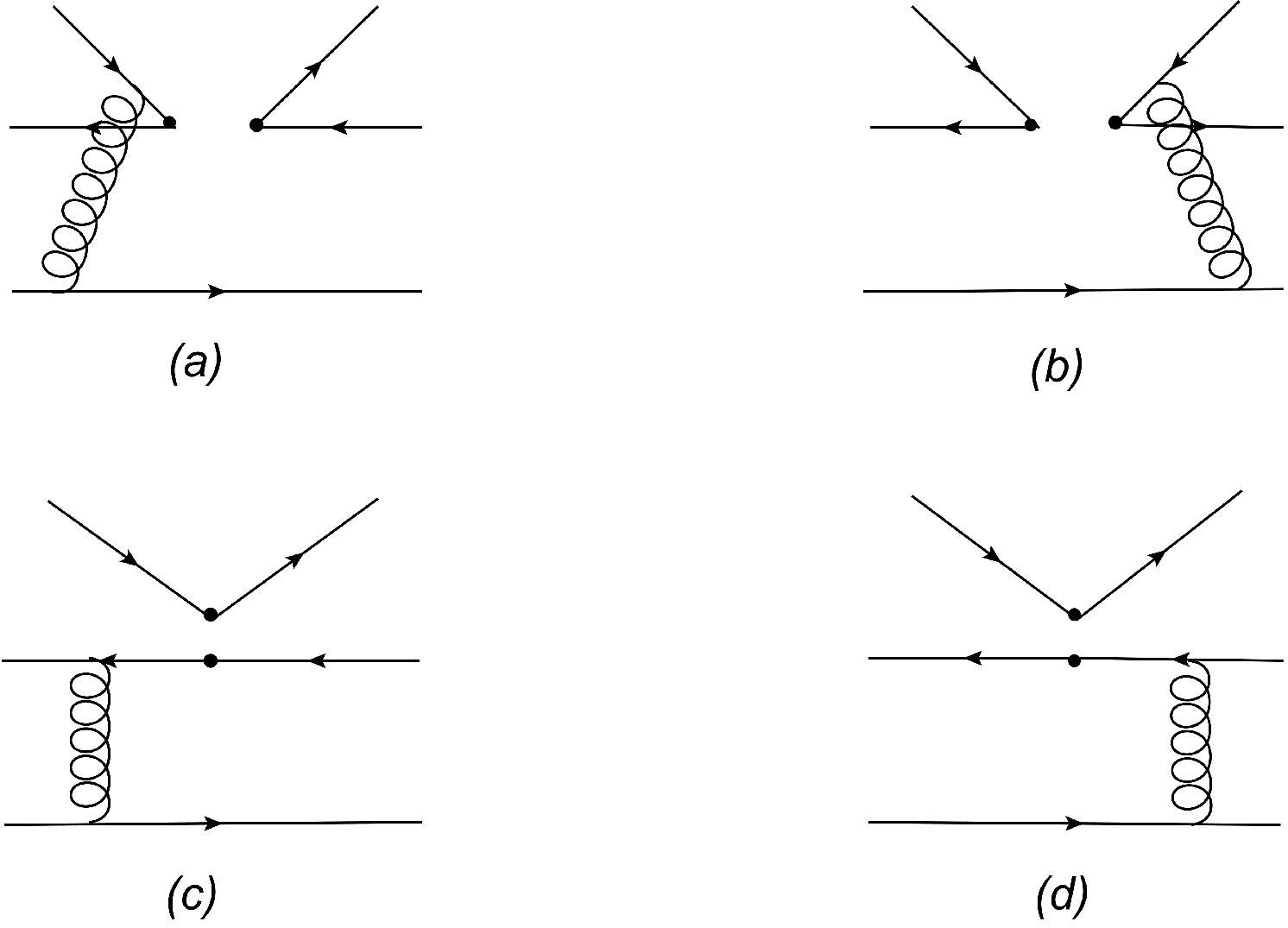}
	\caption{Feynman diagrams for $B_c \to J/\psi A$ decays. (a,b): The nonfactorizable emission diagrams; (c,d): The factorizable emission diagrams.}
	\label{fig:fig1}
\end{figure}

\begin{figure}
	\centering
	\includegraphics[width=0.5\linewidth]{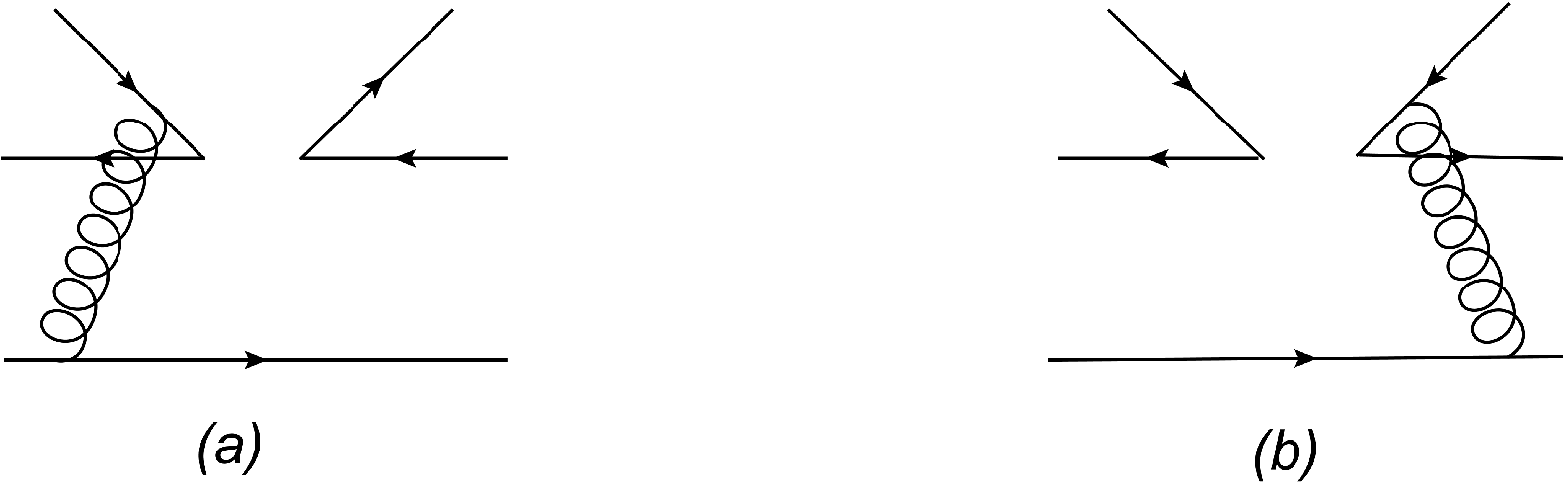}
	\caption{Feynman diagrams for $B_c \to J/\psi T$ decays. (a,b): The nonfactorizable emission diagrams.}
	\label{fig:fig2}
\end{figure}

\begin{equation}
\begin{aligned}
{\cal A}(B_{c} \rightarrow \left(J/\psi A\right)= V_{cb}^{*}V_{ud(s)}\left[\left(C_{2}+\frac{1}{3} C_{1}\right)\mathcal{F}_{e}+C_{1}\mathcal{M}_{e}\right]\label{amplitude A}
\end{aligned}
\end{equation}
here, the Wilson coefficients $C_1$ and $C_2$ are convolved with the amplitudes $\mathcal{F}_e$ and $\mathcal{M}_e$. For the $B_c \to J/\psi(a_1,b_1)$ decay, the amplitude exhibits distinct polarization characteristics, comprising one longitudinal component and two transverse components. The amplitude can be decomposed as:
\begin{equation}
\mathcal{A} = \mathcal{A}^{L} + \mathcal{A}^{N} \epsilon_{2}^{T} \cdot \epsilon_{3}^{T} + i \mathcal{A}^{T} \epsilon_{\alpha\beta\rho\sigma} n^{\alpha} v^{\beta} \epsilon_{2}^{T\rho} \epsilon_{3}^{T\sigma},\label{A}
\end{equation}
here, $\epsilon_{2}^{T}$ and $\epsilon_{3}^{T}$ are the transverse polarization vectors for the vector meson $J/\psi$ and the axial-vector meson ($a_1$ or $b_1$), respectively. $\mathcal{A}^{L}$ corresponds to the longitudinal polarization contribution, while $\mathcal{A}^{N}$ and $\mathcal{A}^{T}$ describe the normal and transverse polarization contributions respectively. The total amplitudes $\mathcal{A}^{L}$, $\mathcal{A}^{N}$, and $\mathcal{A}^{T}$ share the same structure as Eq.\eqref{amplitude A}. The factorization formulas for all polarization components (longitudinal, normal, and transverse) are provided in Appendix.

In the decay $ B_c \to J/\psi \, T $, the tensor meson cannot be directly produced through standard weak operators due to spin constraints, resulting in a vanishing factorizable amplitude. Consequently, the total decay is entirely dominated by non-factorizable emission diagrams in Fig.\ref{fig:fig2}.


\section{Numerical Results And Discussions}\label{sec:numer}

The branching ratios for the decay process of the $B_c$ meson in its rest frame can be expressed as:

\begin{equation}
\mathcal{BR}\left(B_{c} \rightarrow J/\psi \,(A, T)\right) \\
= \frac{G_{F}^{2} \tau_{B_{c}}}{32 \pi M_{B}} \sqrt{1-\left(r_{A(T)}-r_{J / \psi}\right)^{2}} \sqrt{1-\left(r_{A(T)}+r_{J / \psi}\right)^{2}}|\mathcal{A}|^{2},
\end{equation}
where $\tau_{B_c}$ is the lifetime of the $B_c$ meson, and the decay amplitudes $\mathcal{A}$ for each channel have been explicitly given in the appendix. When the final states involve a vector charmonium state and a axial-vector meson or tensor meson, the decay amplitude can be decomposed into three polarized components.

\begin{equation}
\begin{split}
|\mathcal{A}|^{2} = \left| \mathcal{A}_{0} \right|^{2} + \left| \mathcal{A}_{\parallel} \right|^{2} + \left| \mathcal{A}_{\perp} \right|^{2}
\end{split}
\end{equation}
where the amplitudes $ A_0 $, $ A_{\parallel} $, and $ A_{\perp} $ denote the longitudinal, parallel, and perpendicular polarization amplitudes  in the transverses basis, respectively. These relate to the helicity-basis amplitudes $ A^{L} $, $ A^{N}$, and $ A^{T} $ in Eq.(\ref{A}) through a transformation relation.
\begin{equation}
\begin{split}
\mathcal{A}_{0} = \mathcal{A}^{L}, \quad
\mathcal{A}_{\parallel} = \sqrt{2}\mathcal{A}^{N}, \quad
\mathcal{A}_{\perp} = \sqrt{2}\mathcal{A}^{T}
\end{split}
\end{equation}

Turning to the polarizations for $ B_c^+ \to J/\psi (A,T) $ decays, we usually define the polarization fractions
$$
f_{\lambda} = \frac{\left| \mathcal{A}_{\lambda} \right|^{2}}{\left| \mathcal{A}_{0} \right|^{2} + \left| \mathcal{A}_{\parallel} \right|^{2} + \left| \mathcal{A}_{\perp} \right|^{2}}
$$
The polarization fractions satisfy the normalization constraint $\sum_{\lambda} f_{\lambda} = 1$,where $f_0$, $f_{\parallel}$, and $f_{\perp}$ correspond to the longitudinal, parallel, and perpendicular polarization components, respectively.

In Table \ref{tab:params}, we list the input parameters used in the calculations, including meson masses and decay constants, the lifetime of the $B_c$ meson, and the Wolfenstein parameters of the CKM matrix.

\begin{table}[htbp]
	\centering
	\caption{ Various parameters involved in the calculation process}
	\label{tab:params}
	\renewcommand{\arraystretch}{1.6}
	\setlength{\tabcolsep}{12pt}
	\begin{tabular}{lccc}
		\hline\hline
		\\[-5pt]

		Mass of mesons &
		$M_{W}=80.41~\mathrm{GeV}$ &
		$M_{B_c}=6.275~\mathrm{GeV}$ &
		$m_{b}=4.8~\mathrm{GeV}$ \\[3pt]
		&
		$m_{c}=1.5~\mathrm{GeV}$ &
		$m_{J/\psi}=3.097~\mathrm{GeV}$ &
		$m_{a_{1}}=1.23~\mathrm{GeV}$ \\[3pt]
		&
		$m_{b_{1}}=1.21~\mathrm{GeV}$ &
		$m_{a_{2}}=1.317~\mathrm{GeV}$ &
		$m_{K_{2}^{*}(1430)}=1.427~\mathrm{GeV}$   \\[8pt]

		Decay constants of mesons &
		$f_{B_{c}}=489 \pm 4 \pm 3~\mathrm{MeV}$ &
		$f_{J/\psi}=405 \pm 14~\mathrm{MeV}$ &
		$f_{a_{1}}=238~\mathrm{MeV}$ \\[3pt]
		&
		$f_{b_{1}}=180~\mathrm{MeV}$ &
		$f_{a_{2}}=107 \pm 6~\mathrm{MeV}$ &
		$f_{K^{*}_{2}(1430)}=118 \pm 5~\mathrm{MeV}$ \\[8pt]

		Lifetime of meson &
		$\tau_{B_{c}}=0.507~\mathrm{ps}$ &
		& \\[8pt]

		Wolfenstein parameters &
		$A=0.826$ &
		$\lambda=0.22501$ &
		\\[3pt]
		&
		$\bar{\rho}=0.1591$ &
		$\bar{\eta}=0.3523$ &
		\\[5pt]
		
		\hline\hline
	\end{tabular}
\end{table}
Our numerical results for the branching ratios of the $B_c \to J/\psi A$ and $B_c \to J/\psi T$ decay processes are listed in Table~\ref{tab:branching_ratios}. The uncertainties in our calculation results originate from three primary sources:
the first kind stems from hadronic parameters ($\omega_B =( 0.60 \pm 0.05)~\mathrm{GeV}$ \cite{Zhou2012} for $B_c$ meson and $\omega = (0.60 \pm 0.10)~\mathrm{GeV}$ \cite{Rui:2014} for $J/\psi$ meson),
the second kind pertains to decay constants of the $B_c$ meson and final-state charmonium mesons, which can be found in Table \ref{tab:params}
and the third kind arises from $\Lambda_{\mathrm{QCD}} = (0.25 \pm 0.05)~\mathrm{GeV}$ and a $20\%$ variation in the hard scale $t_{\max} = (1.0 \pm 0.2)t$ as defined in the Appendix,
while other uncertainties such as those from CKM matrix elements $V$ related to $\bar{\eta}$ and $\bar{\rho}$ angles of the unitary triangle are found to be negligible.

\begin{table}[htbp]
	\centering
	\caption{Branching ratios for $B_c \to J/\psi\,(A,\,T)$. The errors for these entries correspond respectively to uncertainties in hadronic shape parameters, decay constants, and scale dependence.}
	\label{tab:branching_ratios}
	\renewcommand{\arraystretch}{1.8}
	\setlength{\tabcolsep}{0.8em}
	\begin{tabular}{lcccc}
		\hline\hline
		Decay Modes &
		$B_0$ &
		$B_\parallel$ &
		$B_\perp$ &
		$B_{\text{total}}$ \\
		\hline
		$B_c^+ \to J/\psi a_1^+$ 
		& $7.63^{+1.10+0.44+0.33}_{-0.99-0.42-0.31} \times 10^{-3}$ 
		& $1.75^{+0.26+0.10+0.07}_{-0.22-0.10-0.08} \times 10^{-3}$ 
		& $1.68^{+0.21+0.08+0.07}_{-0.21-0.09-0.06} \times 10^{-3}$ 
		& $11.06^{+1.57+0.62+0.47}_{-1.42-0.61-0.45} \times 10^{-3}$  \\[8pt]
		
		$B_c^+ \to J/\psi b_1^+$ 
		& $1.36^{+0.17+0.11+0.06}_{-0.14-0.09-0.06} \times 10^{-3}$ 
		& $0.09^{+0.01+0.01+0.01}_{-0.02-0.01-0.01} \times 10^{-3}$ 
		& $0.06^{+0.01+0.01+0.00}_{-0.01-0.00-0.00} \times 10^{-3}$ 
		& $1.51^{+0.19+0.13+0.07}_{-0.17-0.10-0.07} \times 10^{-3}$ \\[8pt]
		
		$B_c^+ \to J/\psi a_2^+$ 
		& $2.77^{+0.51+0.16+0.15}_{-0.47-0.12-0.14} \times 10^{-4}$ 
		& $0.07^{+0.10+0.02+0.03}_{-0.04-0.02-0.04} \times 10^{-4}$ 
		& $0.05^{+0.07+0.01+0.02}_{-0.02-0.01-0.02} \times 10^{-4}$ 
		& $2.89^{+0.68+0.19+0.20}_{-0.53-0.15-0.20} \times 10^{-4}$  \\[8pt]
		
		$B_c^+ \to J/\psi K_2^{*+}$ 
		& $4.26^{+0.40+0.04+0.09}_{-0.33-0.04-0.09} \times 10^{-5}$ 
		& $0.12^{+0.03+0.01+0.01}_{-0.03-0.01-0.01} \times 10^{-5}$ 
		& $0.10^{+0.02+0.00+0.00}_{-0.01-0.00-0.00} \times 10^{-5}$ 
		& $4.48^{+0.45+0.05+0.10}_{-0.37-0.05-0.10} \times 10^{-5}$  \\
		\hline\hline
	\end{tabular}
\end{table}

\begin{table}[htbp]
	\centering
	\caption{Polarization fractions  for $B_c^+ \to J/\psi\,(A,\,T)$ decays.
		The errors correspond respectively to uncertainties in hadronic shape parameters, decay constants, and scale dependence.}
	\label{tab:polarization_fractions}
	\renewcommand{\arraystretch}{1.5}
	\setlength{\tabcolsep}{12pt}
	\begin{tabular}{lccc}
		\hline\hline
		Decay Modes &
		$f_0$ &
		$f_\parallel$ &
		$f_\perp$ \\
		\hline
		$B_c^+ \to J/\psi a_1^+$ 
		& $0.69^{+0.02+0.02+0.01}_{-0.02-0.03-0.01}$ 
		& $0.15^{+0.02+0.01+0.00}_{-0.02-0.02-0.00}$ 
		& $0.15^{+0.03+0.02+0.00}_{-0.02-0.02-0.01}$ \\[8pt]
		
		$B_c^+ \to J/\psi b_1^+$ 
		& $0.90^{+0.03+0.03+0.01}_{-0.03-0.02-0.01}$ 
		& $0.06^{+0.01+0.01+0.00}_{-0.01-0.01-0.01}$ 
		& $0.04^{+0.00+0.00+0.00}_{-0.00-0.00-0.00}$ \\[8pt]
		
		$B_c^+ \to J/\psi a_2^+$ 
		& $0.96^{+0.02+0.03+0.00}_{-0.02-0.02-0.01}$ 
		& $0.02^{+0.01+0.01+0.00}_{-0.01-0.00-0.00}$ 
		& $0.02^{+0.00+0.00+0.00}_{-0.00-0.00-0.00}$ \\[8pt]
		
		$B_c^+ \to J/\psi K_2^{*+}$ 
		& $0.95^{+0.02+0.01+0.01}_{-0.01-0.02-0.00}$ 
		& $0.03^{+0.01+0.00+0.00}_{-0.01-0.00-0.00}$ 
		& $0.02^{+0.00+0.01+0.01}_{-0.00-0.00-0.00}$ \\
		\hline\hline
	\end{tabular}
\end{table}

In Table \ref{tab:branching_ratios}, we list the branching ratios for $B_c\to J/\psi A$ and $B_c \to J/\psi T$ decays, where $A$ denotes axial-vector mesons ($a_1$, $b_1$) and $T$ denotes tensor mesons ($a_2$, $K_2^*$). The key observations are:

 (1)For the decay processes $B_c^+ \to J/\psi a_1^+$ and $B_c^+ \to J/\psi b_1^+$, both final states contain axial-vector mesons and exhibit relatively large branching ratios. This is primarily because $a_1^+$ and $b_1^+$ share the same quark composition (mainly $u\bar{d}$), and their decay amplitudes at the tree level benefit from the enhanced CKM matrix element contribution ($|V_{cb}V_{ud}^{*}| \sim 10^{-2}$). In Table \ref{tab:branching_ratios}, we can see:
 $$\mathcal{B}(B_c^+ \to J/\psi b_1^+) < \mathcal{B}(B_c^+ \to J/\psi a_1^+)$$ 
 The key reason lies in the effective decay constant of the $b_1^+$ meson, $f_{b_{1}^{+}} = f_{b_{1}} \cdot a_{0 b_{1}}^{\parallel} \sim 0.0005$, which is nearly vanishing~\cite{Liu:2023bc}. This strongly suppresses the factorizable emission diagram contributions that depend on this constant. As a result, the branching ratio of $B_c^+ \to J/\psi b_1^+$ arises almost entirely from non-factorizable emission decay amplitudes.
Nevertheless, the $b_1(1235)$, as a ${}^1P_1$ state, possesses a leading-twist distribution amplitude with odd-parity antisymmetric behavior. This symmetry characteristic alters the quantum interference pattern between specific Feynman diagrams (e.g., those involving quark-antiquark momentum exchange) in the non-factorizable emission topology: in $B_c^+ \to J/\psi a_1^+$ (where $a_1$ is a ${}^3P_1$ state with a symmetric wave function), the non-factorizable diagrams exhibit destructive interference; while in $B_c^+ \to J/\psi b_1^+$, the same non-factorizable topology undergoes constructive interference due to the antisymmetric wave function.
 This crucial interference transition effectively compensates for the suppression caused by the extremely small decay constant, allowing the branching ratio of the $b_1^+$ channel to remain at the order of $10^{-3}$.

(2) To facilitate future experimental investigations, we define the ratios of branching fractions using $\mathcal{B}(B_c^+ \to J/\psi \pi^+)$ from Ref.~\cite{Rui:2014} as a benchmark. The benchmark decay $B_c^+ \to J/\psi \pi^+$ is chosen because it is theoretically clean and provides a well-normalized reference within the factorization approach. The calculated ratios are as follows:
\begin{equation}
	\begin{split}
	R_{a_1/\pi} &\equiv \frac{\mathcal{B}(B_c^+ \to J/\psi a_1^+)}{\mathcal{B}(B_c^+ \to J/\psi \pi^+)} \approx 4.75,\\[4pt]
	R_{b_1/\pi} &\equiv \frac{\mathcal{B}(B_c^+ \to J/\psi b_1^+)}{\mathcal{B}(B_c^+ \to J/\psi \pi^+)} \approx 0.65,\\[4pt]
	R_{a_2/\pi} &\equiv \frac{\mathcal{B}(B_c^+ \to J/\psi a_2^+)}{\mathcal{B}(B_c^+ \to J/\psi \pi^+)}  \approx 0.124,\\[4pt]
	R_{K_2^{*}/\pi} &\equiv \frac{\mathcal{B}(B_c^+ \to J/\psi K_2^{*+})}{\mathcal{B}(B_c^+ \to J/\psi \pi^+)}  \approx 0.019.
	\end{split}
\end{equation}
These four ratios reveal a distinct hierarchy:
$$
R_{a_1/\pi}\;\gg\; R_{b_1/\pi}  \;>\; R_{a_2/\pi} \;\gg\; R_{K_2^{*}/\pi} 
$$
The large value of $R_{a_1/\pi}$ reflects the dominance of factorizable contributions, enhanced by the sizable decay constant of the axial-vector $a_1^+$ meson. In contrast, the significantly smaller $R_{b_1/\pi}$ highlights the suppression associated with non-factorizable dynamics and the peculiar $J^{PC}=1^{+-}$ quantum number of the $b_1^+$ meson. The even smaller ratios for the tensor mesons, $R_{a_2/\pi}$ and $R_{K_2^{*}/\pi}$, indicate further suppression in decays to higher-spin ($J=2$) final states, which involve more intricate helicity structures and receive contributions from subleading power corrections. Notably, the strange tensor meson $K_2^{*+}$ exhibits an order-of-magnitude further suppression relative to the non-strange $a_2^+$, a pattern attributable to the Cabibbo-suppressed $b \to s$ transition and to possible differences between the distribution amplitudes of strange and non-strange tensor mesons.
These ratios provide a sensitive set of observables for testing the interference between color-favored and color-suppressed amplitudes, the role of non-factorizable dynamics, and the spin-dependent hadronization mechanisms in $B_c$ decays. Precise future measurements of these channels are important for several reasons: first, they offer a direct test of pQCD frameworks in describing non-factorizable contributions in processes involving heavy quarkonia; second, comparison with theoretical predictions will help to constrain hadronic parameters, especially the distribution amplitudes of the $B_c$ meson and the light mesons; and finally, any substantial discrepancy between experiment and theory could signal physics beyond the Standard Model, making these decays a valuable platform for indirect searches for new physics. The inclusion of the tensor meson channels further enriches this program by probing spin-dependent dynamics and the factorization properties of higher-twist distribution amplitudes.

(3)A comparison with existing literature highlights significant refinements in both the theoretical framework and input parameters employed in this study. For the decay $B_c^+ \to J/\psi a_1(1260)^+$, the branching ratio reported in Ref.\cite{Najafi2015} is 
$\left(1.02_{-0.08-0.05-0.01}^{+0.04+0.03+0.01}\right) \times 10^{-3}$, which differs from our computed value for the same channel. These discrepancies stem primarily from two sources. First, at the level of theoretical framework, we have incorporated an intrinsic transverse-momentum $b$-dependence into the wave function model of the $B_c$ meson, as given in Eq.~\eqref{PhiBc}. This improvement allows a more accurate description of transverse-momentum effects in the hadronic wave function, leading to enhanced precision in handling both factorizable and non-factorizable diagrams, especially those involving soft-gluon exchange, thereby reducing systematic uncertainties originating from model simplifications.
Second, regarding input parameters, we have updated the CKM matrix elements, meson masses, decay constants, and hadronic shape parameters in accordance with the latest experimental data and theoretical developments. This update ensures that our predictions are based on the most reliable available inputs, making the results not only a test of the pQCD framework but also a timely and competitive theoretical reference for future experimental measurements.Therefore, the observed differences signify a substantive evolution in the theoretical methodology. This methodological advancement, utilizing an improved wave function and contemporary parameters, yields more reliable predictions. These results provide a refined benchmark for high-precision experiments at the upgraded LHCb and will deepen the understanding of heavy-flavored hadron decay dynamics.

(4) Compared with Ref.\cite{Liu:2023bc}, the branching-ratio predictions for $B_c^+ \to J/\psi A$ and $B_c^+ \to J/\psi T$ in the present work exhibit differences, yet the results of both approaches lie within the same order of magnitude. This consistency in the overall scale indicates that, despite deviations due to distinct theoretical treatments, the two methods yield compatible estimates of the basic branching ratio features for $B_c$ decays, thereby reinforcing the robustness of the theoretical predictions.
The numerical differences mainly stem from the fundamental distinction between the two theoretical frameworks. In Ref.\cite{Liu:2023bc}, a modified Sudakov factor $s_c(Q,b)$ is introduced to explicitly incorporate the charm-quark mass scale $m_c$ into the evolution kernel, thereby addressing radiative corrections associated with the ratio $m_c/m_b$. In contrast, the present study employs a $B_c$ meson wave function of the form $\delta(x - m_c/M_{B_c})$, which fixes the longitudinal momentum fraction of the charm quark and naturally embeds the double-heavy flavor nature of the $B_c$ meson at the level of the initial-state wave function. Theoretically, the two schemes improve the description accuracy from complementary perspectives: the $\delta$-type wave function adopted here reduces the sensitivity to the endpoint behavior ($x \to 0,1$) of the light-meson distribution amplitudes by locking the parton momentum fraction, while the improved Sudakov factor $s_c(Q,b)$ in Ref.\cite{Liu:2023bc} explicitly includes the $m_c$ contribution in the resummation, allowing for a more precise treatment of soft-gluon radiation at the charm-quark mass scale. These complementary treatments enhance the descriptive power of the theoretical framework for the double-heavy $B_c$ system from the angles of the initial-state wave function and the evolution kernel, respectively.
Given that the two methods emphasize different aspects of the theoretical treatment, the present work and Ref.\cite{Liu:2023bc} together constitute two complementary realizations of the pQCD approach in $B_c$ physics. The cross-validation of their theoretical architectures and numerical outcomes provides a more comprehensive and reliable theoretical reference for future precision measurements at experiments such as LHCb and Belle II, and extends our understanding of non-perturbative QCD effects at the charm-quark mass scale from different perspectives.

(5) To quantitatively probe the distinct decay dynamics between the vector $\rho$ meson and the axial-vector $a_1$ and $b_1$ mesons, we define the following two ratios, normalized to the theoretically established branching fraction of the $B_c^+ \to J/\psi \rho^+$ decay as a reference~\cite{Rui:2014}: 
\begin{equation}
\begin{split}
	R_{a_1/\rho} \equiv \frac{\mathcal{B}(B_c^+ \to J/\psi a_1^+)}{\mathcal{B}(B_c^+ \to J/\psi \rho^+)}\approx1.35, \\
	R_{b_1/\rho} \equiv \frac{\mathcal{B}(B_c^+ \to J/\psi b_1^+)}{\mathcal{B}(B_c^+ \to J/\psi \rho^+)}\approx0.18.	
	\end{split}
\end{equation}
The $B_c^+ \to J/\psi \rho^+$ channel serves as a well-understood benchmark within the factorization framework, being dominated by factorizable emission diagrams with clear vector-current contributions. This choice provides a theoretically clean and reliable baseline, enabling a quantitative comparison and interpretation of the more intricate dynamics governing the decays into axial-vector final states.
In the Standard Model, the axial-vector $a_1$ meson and the vector meson $\rho$ exhibit similar QCD hadronic dynamics~\cite{Cheng:2008}. Hence, the decay pattern of $B_c^+ \to J/\psi a_1^+$ is expected to resemble that of $B_c^+ \to J/\psi \rho^+$. The ratio $R_{a_1/\rho}$ reveals that the branching ratio $\mathcal{B}(B_c^+ \to J/\psi a_1^+)$ is approximately 35\% larger than $\mathcal{B}(B_c^+ \to J/\psi \rho^+)$. This enhancement reflects the dominance of factorizable contributions in the $a_1$ channel, likely attributable to its larger decay constant and similar vector-current-driven dynamics. In contrast, $R_{b_1/\rho}$ indicates that $\mathcal{B}(B_c^+ \to J/\psi b_1^+)$ is suppressed by about 82\% relative to $\mathcal{B}(B_c^+ \to J/\psi \rho^+)$. This significant suppression originates from the strong suppression of factorizable amplitudes in $b_1$ production, underscoring the importance of non-factorizable contributions and the distinctive internal structure of the $b_1$ meson ($J^{PC}=1^{+-}$). The marked difference between $R_{a_1/\rho}$ and $R_{b_1/\rho}$ highlights the sensitivity of these ratios to the internal spin-parity configurations of axial-vector mesons, providing a clear experimental signature for distinguishing $1^{++}$  and $1^{+-}$  states.
These ratios constitute powerful probes for testing the underlying dynamics of heavy-flavor decays both within and beyond the Standard Model. The enhanced $R_{a_1/\rho}$ validates the dominance of factorizable mechanisms in $a_1$ production, offering a benchmark for pQCD predictions, while the suppressed $R_{b_1/\rho}$ emphasizes the necessity of incorporating non-factorizable effects that are sensitive to QCD vacuum structures and higher-order corrections. Future high-precision experiments at facilities such as the upgraded LHCb, Belle II, or a future Electron-Ion Collider can leverage these observables to constrain hadronic parameters like decay constants and distribution amplitudes, discriminate between different theoretical factorization schemes, probe possible exotic configurations such as hybrid or tetraquark admixtures in axial-vector mesons, and search for signatures of new physics through deviations from Standard Model predictions. Consequently, precise measurements of $R_{a_1/\rho}$ and $R_{b_1/\rho}$ will not only deepen our understanding of non-perturbative QCD but also significantly enhance the sensitivity of $B_c$ decays to potential new physics scenarios.

(6) For the decays $B_c^+ \to J/\psi a_2^+$ and $B_c^+ \to J/\psi K_2^{*+}$, the factorizable emission diagrams vanish due to a fundamental mismatch in quantum numbers: tensor mesons ($J^{PC} = 2^{++}$) cannot be directly created from the vacuum via local $(V \pm A)$ or $(S \pm P)$ weak currents, which only generate states with $J \leq 1$. Consequently, $\langle T \vert \bar{q} \Gamma q \vert 0 \rangle = 0$, forbidding the factorizable emission process in the Standard Model. These decays therefore proceed entirely through the non-factorizable emission mechanism, which leads to their characteristically small branching fractions.
Furthermore, the comparison between the $a_2^+$ channel (involving a $b \to d$ transition) and the $K_2^{*+}$ channel (involving a $b \to s$ transition) serves as a sensitive probe of flavor-SU(3) symmetry breaking. The predicted ratio of their branching fractions exhibits a significant deviation from the expectation based solely on CKM suppression, highlighting the influence of differences in non-perturbative hadronic parameters, such as decay constants and distribution amplitudes, between strange and non-strange tensor mesons.
The study of these decays thus provides a unique laboratory to test QCD dynamics in heavy-flavor decays, offering critical insights into the properties of tensor mesons and the breaking of flavor symmetries. These investigations not only validate the capability of theoretical frameworks like pQCD in describing exclusively non-factorizable topologies but also substantially advance our understanding of strong interaction mechanisms.

Turning to Table \ref{tab:polarization_fractions}, we present a systematic analysis of the predicted polarization fractions of mesons:
It can be see that the all decays  reveal profound insights into heavy-quark hadronization dynamics. Three distinct polarization patterns emerge with striking consistency across decay channels, reflecting the universal dominance of longitudinal polarizations ($ f_0 $) in branching ratios. In particular, the highest longitudinal polarization fractions observed for the tensor meson final states $B_c^+ \to J/\psi a_2^+$ and $B_c^+ \to J/\psi K_2^{*+}$, which nearly approach complete longitudinal polarization. This behavior stems from the tensor mesons' orbital angular momentum of $L=1$ (corresponding to the quantum number assignment $J^{PC} = 2^{++}$). Specifically, the endpoint contributions in their LCDAs substantially enhance the convolution weight of the longitudinal helicity component. Concurrently, the mismatch in quantum numbers forbids the contributions from factorizable emission diagrams, confining these decays to proceed entirely via non-factorizable mechanisms and thus further suppressing transverse polarization.
In contrast, the longitudinal polarization fraction for $B_c^+ \to J/\psi a_1^+$ is notably lower. This is attributed to the large decay constant of $a_1^+$ as a $^3P_1$ state, which sustains considerable factorizable emission contributions that generally favor transverse polarization to a certain extent. It is worth emphasizing that the longitudinal polarization fraction for $B_c^+ \to J/\psi b_1^+$ is significantly higher than that of the $a_1^+$ channel. The primary reason lies in the nearly vanishing effective decay constant of $b_1^+$, a feature arising from its $^1P_1$ quantum numbers. This tiny decay constant strongly suppresses any factorizable diagrams dependent on this parameter, forcing the decay to be dominated entirely by non-factorizable contributions. Furthermore, the antisymmetric distribution amplitude of the $b_1$ meson transforms the quantum interference in non-factorizable diagrams from destructive (as observed in the $a_1^+$ channel) to constructive, which further amplifies the longitudinal polarization amplitude.
Regarding transverse polarization, both the parallel ($f_{\parallel}$) and perpendicular ($f_{\perp}$) components are drastically suppressed across all decay channels. This suppression is particularly pronounced in tensor meson decays, where the transverse polarization components are reduced to an extremely low level. This phenomenon is consistent with the helicity selection rules for the $0^- \to 1^-$ effective transition of the heavy quark line, under the approximation of heavy-quark helicity conservation (HQHC). Moreover, the polarization fractions of the strange ($K_2^{*+}$) and non-strange ($a_2^+$) tensor meson channels are in excellent agreement within uncertainties. This indicates that polarization is relatively insensitive to the flavor composition of the light quarks. Nevertheless, the minor discrepancies in $f_{\parallel}$ and $f_{\perp}$ may reflect the distortions of LCDAs induced by the non-vanishing mass of the strange quark.
These polarization patterns not only verify the capability of the pQCD framework to describe the helicity structure of double-heavy meson systems, but also provide clear theoretical benchmarks. These benchmarks will facilitate the discrimination of the intrinsic angular momentum structures of different mesons through angular distribution measurements in future experiments at the LHCb and Belle II collaborations.

\section{Summary} \label{sec:summary}

In this article, we systematically investigate the exclusive two-body decays of the $B_c$ meson to an $S$-wave charmonium state ($J/\psi$) and either axial-vector mesons ($a_1$, $b_1$) or tensor mesons ($a_2$, $K_2^*$) within the pQCD factorization framework, incorporating the nonperturbative distribution amplitudes of the final-state mesons to ensure infrared safety. Since flavor non-degeneracy in four-quark operators forbids penguin diagrams, weak phase differences vanish, eliminating CP violation in these decays.
We focus on computing the branching ratios and polarization fractions for the aforementioned decay channels. To provide sensitive observables for experimental verification, this study defines ratios of branching fractions using the benchmark decays $B_c^+ \to J/\psi \pi^+$ and $B_c^+ \to J/\psi \rho^+$ \cite{Rui:2014}, which are well-studied channels with established theoretical and experimental results. These ratios not only quantify the relative significance of factorizable versus non-factorizable contributions but also encode spin-dependent hadronization effects and flavor-$SU(3)$ symmetry breaking. Furthermore, they serve as stringent probes for testing the QCD factorization framework and constraining key non-perturbative hadronic parameters, which are pivotal for refining the pQCD formalism in heavy-flavor physics. The obtained results provide a solid theoretical basis for experimental investigations and are particularly well-suited for leveraging the high-statistics environment of the LHCb experiment.
This work not only validates the applicability of the pQCD framework in describing the helicity structure and strong interaction dynamics of double-heavy mesons but also delivers valuable theoretical predictions for ongoing and future studies at the LHCb and Belle II experiments. These findings will further deepen our understanding of non-perturbative QCD effects, meson internal structure, and flavor symmetry breaking mechanisms in heavy-flavor decays, while laying the groundwork for the indirect detection of new physics signals.


\section*{acknowledgments}

This work is supported by the National Natural Science Foundation of China under Grant No.11047028.

\section*{Appendix : FACTORIZATION FORMULAE} \label{sec:appendix}
\appendix
\setcounter{equation}{0}
\renewcommand\theequation{A.\arabic{equation}}
Similar to vector mesons, axial-vector mesons possess intrinsic spin degrees of freedom, which necessitates distinct amplitude components for the three polarization states (longitudinal, normal, transverse) in $ B_c \to J/\psi A $ decays, denoted respectively as $ L $, $ N $ and $ T $ amplitudes.

\begin{eqnarray}
\begin{split}
F_{e}^{L} &= 2\sqrt{\frac{2}{3}}\pi C_{F}f_{B}f_{A} M_{B}^{4} \sqrt{1-r_{2}^{2}} \int_{0}^{1}dx_{2}\int_{0}^{\infty}b_{1}b_{2}db_{1}db_{2} \exp\left(-\frac{\omega_{B}^{2}b_{1}^{2}}{2}\right) \\
&\quad \times\Big\{ \big[r_{2}\phi^{t}(x_{2},b_{2})(r_{b}-2x_{2}) + \phi^{L}(x_{2},b_{2})(x_{2}-2r_{b})\big] E_{ab}(t_{a})h(\alpha_{e},\beta_{a},b_{1},b_{2})S_{t}(x_{2}) \\
&\quad - \phi^{L}(x_{2},b_{2})\big[r_{2}^{2}+r_{c}\big]E_{ab}(t_{b})h(\alpha_{e},\beta_{b},b_{2},b_{1})S_{t}(x_{1}) \Big\}
\end{split}
\end{eqnarray}

\begin{eqnarray}
\begin{split}
M_{e}^{L} &= \frac{8}{3}\pi C_{F}f_{B}M_{B}^{4}\sqrt{1-r_{2}^{2}} \int_{0}^{1}dx_{2}dx_{3}\int_{0}^{\infty}b_{1}b_{3}db_{1}db_{3}\exp\left(-\frac{\omega_{B}^{2}b_{1}^{2}}{2}\right)\phi_{A}(x_{3}) \\
&\quad \times \Big\{\big[-\phi^{L}(x_{2},b_{1})(r_{2}^{2}-1)(x_{1}-x_{3})+ r_{2}\phi^{t}(x_{2},b_{1})(x_{1}+x_{2}-1)\big] E_{cd}(t_{c})h(\beta_{c},\alpha_{e},b_{3},b_{1}) \\
&\quad -\big[\phi^{L}(x_{2},b_{1})(r_{2}^{2}(x_{2}-x_{3})+2x_{1}+x_{2}+x_{3}-2)\\
&\quad - r_{2}\phi^{t}(x_{2},b_{1})(x_{1}+x_{2}-1)\big]E_{cd}(t_{d})h(\beta_{d},\alpha_{e},b_{3},b_{1})\Big\}
\end{split}
\end{eqnarray}

\begin{eqnarray}
\begin{split}
F_{e}^{N} &= 2\sqrt{\frac{2}{3}}\pi C_{F}f_{B}f_{A} M_{B}^{4}r_{3} \int_{0}^{1}dx_{2} \int_{0}^{\infty}b_{1}b_{2}db_{1}db_{2}  \exp\left(-\frac{\omega_{B}^{2}b_{1}^{2}}{2}\right)\\ &\quad\times\Big\{\big[r_{2}\phi^{T}(x_{2},b_{2})(-4r_{b}+x_{2}+1)
 + (r_{b}-2)\phi^{V}(x_{2},b_{2})\big] E_{ab}(t_{a})h(\alpha_{e},\beta_{b},b_{2},b_{1})S_{t}(x_{2})\big]\\
 &\quad+\phi^{T}(x_{2},b_{2})\big[r_{2}(x_{1}-1)\big]E_{ab}(t_{b})h(\alpha_{e},\beta_{b},b_{2},b_{1})S_{t}(x_{1})\big]\Big\}
\end{split}
\end{eqnarray}

\begin{eqnarray}
\begin{split}
M_{e}^{N} &= \frac{8}{3}\pi C_{F}f_{B}M_{B}^{4}r_{3} \int_{0}^{1}dx_{2}dx_{3} \int_{0}^{\infty}b_{1}b_{3}db_{1}db_{3}
\exp\left(-\frac{\omega_{B}^{2}b_{1}^{2}}{2}\right)\\ &\quad \times\Big\{\big[\phi_{A}^{a}(x_{3})\phi^{V}(x_{2},b_{1})(r_{2}^{2}-1)(x_{2}+x_{3}-1)
+ \phi_{A}^{v}(x_{3}) \phi^{T}(x_{2}, b_{1}) \\ &\quad \times\left(r_{2}^{2} (x_{1} + 2x_{2} + x_{3} - 2) + x_{1} - x_{3} \right) \big] E_{cd}(t_{c}) h(\beta_{c}, \alpha_{e}, b_{3}, b_{1})\\ &\quad
+ \big[ \phi_{A}^{a}(x_{3}) 2(r_{2}^{2} - 1) \Big( r_{2} (x_{2} - x_{3}) \phi^{V}(x_{2}, b_{1})  + 2(x_{1} + x_{3} - 1) \phi^{T}(x_{2}, b_{1}) \Big)\\ &\quad
+ \phi_{A}^{v}(x_{3}) ( 2 r_{2} \phi^{V}(x_{2}, b_{1}) \Big( r_{2}^{2} (x_{2} - x_{3}) + 2x_{1} + x_{2} + x_{3} - 2 \Big) \\
&\quad + \phi^{T}(x_{2}, b_{1}) ( r_{2}^{2} (1 - x_{1} - 2x_{2} + x_{3}) + 1 - x_{1} - x_{3} ) ) \big] E_{cd}(t_{d}) h(\beta_{d}, \alpha_{e}, b_{3}, b_{1}) \Big\}
\end{split}
\end{eqnarray}

\begin{eqnarray}
\begin{split}
F_{e}^{T} &= 2\sqrt{\frac{2}{3}}\pi C_{F}f_{B}f_{A}M_{B}^{4}r_{3} \int_{0}^{1}dx_{2} \int_{0}^{\infty}b_{1}b_{2}db_{1}db_{2} \exp\left(-\frac{\omega_{B}^{2}b_{1}^{2}}{2}\right)\\ &\quad \times\Big\{\big[\phi^{V}(x_{2},b_{2})(2-r_{b}) +r_{2}\phi^{T}(x_{2},b_{2})(x_{2}-1)\big]E_{ab}(t_{a})h(\alpha_{e},\beta_{a},b_{1},b_{2})S_{t}(x_{2}) \\
&\quad + \psi^{T}(x_{2},b_{2})[r_{2}(x_{1}-1)]E_{ab}(t_{b})h(\alpha_{e},\beta_{b},b_{2},b_{1})S_{t}(x_{1})\Big\}
\end{split}
\end{eqnarray}

\begin{eqnarray}
\begin{split}
M_{e}^{T} &= \frac{8}{3}\pi C_{F}f_{B}M_{B}^{4}r_{3} \int_{0}^{1}dx_{2}dx_{3} \int_{0}^{\infty} b_{1}b_{3} db_{1} db_{3}  \exp\left(-\frac{\omega_{B}^{2}b_{1}^{2}}{2}\right) \\
&\quad \times\Big\{ \big[-\phi_{A}^{a}(x_{3}) \phi^{V}(x_{2}, b_{1})  2 r_{2}  ( r_{2}^{2} (x_{2} + x_{3} - 1) + 2x_{1} + x_{2} - x_{3} - 1 ) \\&\quad-\phi_{A}^{v}(x_{3}) \phi^{T}(x_{2}, b_{1}) ( r_{2}^{2} - 1 ) (x_{1} - x_{3}) \big]  E_{cd}(t_{c}) h(\beta_{c}, \alpha_{e}, b_{3}, b_{1})\\&\quad -\big[ 2\phi_{A}^{a}(x_{3}) \big( r_{2} \phi^{V}(x_{2}, b_{1})  ( r_{2}^{2} (x_{2} - x_{3}) + 2x_{1} + x_{2} + x_{3} - 2) \\
&\quad + 2\phi^{T}(x_{2}, b_{1}) ( r_{2}^{2} (1 - x_{1} - 2x_{2} + x_{3}) + 1 - x_{1} - x_{3} ) \big) \\
&\quad + \phi_{A}^{v}(x_{3}) \phi^{T}(x_{2}, b_{1}) \left( r_{2}^{2} - 1 \right) (x_{1} + x_{3} - 1) \big]  E_{cd}(t_{d}) h(\beta_{d}, \alpha_{e}, b_{3}, b_{1}) \Big\}
\end{split}
\end{eqnarray}

For the $B_c \to J/\psi T$ decay, the calculation of its amplitude can be mapped to that of the corresponding axial-vector case via the following substitution:
\begin{equation}
\begin{split}
\mathcal{M}_{T}^{L} &= \sqrt{\frac{2}{3}} \mathcal{M}_{A}^{L} \bigg|_{\scriptsize\begin{array}{l}
	\phi_{A}^{L} \to \phi_{T}, \\
	\phi_{A}^{t} \to \phi_{T}^{t}, \\
	r_{c} \to -r_{c}
	\end{array}} \\
\mathcal{M}_{T}^{N,T} &= \sqrt{\frac{1}{2}} \mathcal{M}_{A}^{N,T} \bigg|_{\scriptsize\begin{array}{l}
	\phi_{A}^{T} \to \phi_{T}^{T}, \\
    \phi_{A}^{V} \to \phi_{T}^{V}, \\
	r_{c} \to -r_{c}
	\end{array}}
\end{split}
\end{equation}
The numerical prefactors $\sqrt{\frac{2}{3}}$ and $\sqrt{\frac{1}{2}}$ are derived from the equivalent polarization vector $\epsilon_{\bullet}$ defined in Eq.~\eqref{polarization vector} for the tensor mesons, corresponding to the longitudinal and transverse polarizations, respectively.

The hard functions $ h $ are defined via Fourier transforms of virtual quark and gluon propagators, with specific expressions given by:

\begin{eqnarray}
\begin{split}
h(\alpha, \beta, b_{1}, b_{2})&=h_{1}(\alpha, b_{1})\times h_{2}(\beta, b_{1}, b_{2}),\\
h_{1}(\alpha, b_{1})&=
\begin{cases}
K_{0}(\sqrt{\alpha}b_{1}),&\alpha\textgreater 0,\\
K_{0}(i\sqrt{-\alpha}b_{1}),&\alpha\textless 0,
\end{cases}\\
h_{2}(\beta, b_{1}, b_{2})&=
\begin{cases}
\theta(b_{1}-b_{2})I_{0}(\sqrt{\beta}b_{2})K_{0}(\sqrt{\beta}b_{1})+(b_{1}\leftrightarrow b_{2}),&\beta\textgreater 0,\\
\theta(b_{1}-b_{2})J_{0}(\sqrt{-\beta}b_{2})K_{0}(i\sqrt{-\beta}b_{1})+(b_{1}\leftrightarrow b_{2}),&\beta\textless 0,
\end{cases}
\end{split}
\end{eqnarray}
With $ J_0 $ denoting the Bessel function of the first kind, and $ K_0 $, $ I_0 $ representing the modified Bessel functions of the second and first kinds, respectively, the function $ K_0 $ for a real argument $ x $ is related to $ J_0 $ and the Neumann function $ N_0 $ by $K_0(ix) = \frac{\pi}{2} \left( -N_0(x) + i J_0(x) \right)$. The expressions for $\alpha$ and $\beta$ in the hard functions are

\begin{eqnarray}
\begin{split}
\alpha_{e} &= \left[x_{1} + r_{2}^{2}\left(x_{2} - 1\right)\right]\left[x_{1} + x_{2} - 1 + r_{3}^{2}\left(1 - x_{2}\right)\right] M_{B}^{2}, \\
\beta_{a} &= \left[r_{b}^{2} + \left(1 + r_{2}^{2}\left(x_{2} - 1\right)\right)\left(x_{2} - r_{3}^{2}\left(x_{2} - 1\right)\right)\right] M_{B}^{2}, \\
\beta_{b} &= \left[r_{c}^{2} + \left(r_{2}^{2} - x_{1}\right)\left(x_{1} - 1 + r_{3}^{2}\right)\right] M_{B}^{2}, \\
\beta_{c} &= \left[x_{1} + x_{2} - 1 + r_{3}^{2}\left(1 - x_{2} - x_{3}\right)\right]\left[x_{3} - x_{1} - r_{2}^{2}\left(x_{2} + x_{3} - 1\right)\right] M_{B}^{2}, \\
\beta_{d} &= \left[x_{1} + x_{2} - 1 - r_{3}^{2}\left(x_{2} - x_{3}\right)\right]\left[1 - x_{1} - x_{3} - r_{2}^{2}\left(x_{2} - x_{3}\right)\right] M_{B}^{2}.
\end{split}
\end{eqnarray}

The hard scales $t_{i}(i=a,b,c,d)$, chosen to eliminate large logarithmic radiative corrections in the pQCD factorization framework, are explicitly defined by the following kinematic variables:

\begin{eqnarray}
\begin{split}
t_{a} &= \max \left( \sqrt{\left| \alpha_{e} \right|}, \sqrt{\left| \beta_{a} \right|}, 1/{b_{1}}, 1/{b_{2}} \right),
& t_{b} &= \max \left( \sqrt{\left| \alpha_{e} \right|}, \sqrt{\left| \beta_{b} \right|}, 1/{b_{1}}, 1/{b_{2}} \right), \\
t_{c} &= \max \left( \sqrt{\left| \alpha_{e} \right|}, \sqrt{\left| \beta_{c} \right|}, 1/{b_{1}}, 1/{b_{3}} \right),
& t_{d} &= \max \left( \sqrt{\left| \alpha_{e} \right|}, \sqrt{\left| \beta_{d} \right|}, 1/{b_{1}}, 1/{b_{3}} \right),
\end{split}
\end{eqnarray}

The function $ E_{ij}(t) $ is defined as the product of the running coupling strength $ \alpha_s(t) $ and the Sudakov form factor $ S_{ab,cd}(t) $, which resums the soft gluon radiation contributions:
\begin{equation}
E_{ab,cd}(t) = \alpha_s(t) \, S_{ab,cd}(t)
\end{equation}
where the Sudakov exponents are defined as

\begin{eqnarray}
\begin{split}
S_{ab}(t) &= s\left( \frac{M_B}{\sqrt{2}} x_1, b_1 \right) + s\left( \frac{M_B}{\sqrt{2}} x_2, b_2 \right) + s\left( \frac{M_B}{\sqrt{2}} (1 - x_2), b_2 \right) + \frac{5}{3} \int_{1/b_1}^{t} \frac{d\mu}{\mu} \gamma_q(\mu) + 2 \int_{1/b_2}^{t} \frac{d\mu}{\mu} \gamma_q(\mu), \\
S_{cd}(t) &= s\left( \frac{M_B}{\sqrt{2}} x_1, b_1 \right) + s\left( \frac{M_B}{\sqrt{2}} x_2, b_1 \right) + s\left( \frac{M_B}{\sqrt{2}} (1 - x_2), b_1 \right)+ s\left( \frac{M_B}{\sqrt{2}} x_3, b_3 \right) + s\left( \frac{M_B}{\sqrt{2}} (1 - x_3), b_3 \right) \\
&\quad + \frac{11}{3} \int_{1/b_1}^{t} \frac{d\mu}{\mu} \gamma_q(\mu) + 2 \int_{1/b_3}^{t} \frac{d\mu}{\mu} \gamma_q(\mu),
\end{split}
\end{eqnarray}
 where the quark field anomalous dimension $ \gamma_q = -\alpha_s / \pi $ governs its renormalization group evolution at leading order, with $ \alpha_s $ the strong coupling constant. The Sudakov factor $ s(Q,b) $ can be found in Ref.~\cite{Cheng2006}


\end{document}